\documentclass[seceq,supplement]{ptptex}

\usepackage{graphicx}
\usepackage{wrapft}

\newcommand{\alt}{\mathrel{\raise.3ex\hbox{$<$\kern
-.75em\lower1ex\hbox{$\sim$}}}}
\newcommand{\agt}{\mathrel{\raise.3ex\hbox{$>$\kern
-.75em\lower1ex\hbox{$\sim$}}}}
\newcommand{\tens}{\boldsymbol}
\renewcommand{\vec}{\boldsymbol}
\newcommand{\imag}{i}
\newcommand{\eul}{e}

\title{
Density-matrix renormalization group methods for momentum- and 
frequency-resolved dynamical correlation functions%
}

\author{
Eric \textsc{Jeckelmann}%
}

\inst{
Institut f\"{u}r Theoretische Physik, Leibniz Universit\"{a}t Hannover,
Appelstra\ss e 2, 30167 Hannover, Germany  
}

\abst{
Several density-matrix renormalization group methods
have been proposed to compute the momentum- and frequency-resolved dynamical 
correlation functions
of low-dimensional strongly correlated systems.
The most relevant approaches are discussed in this contribution.
Their applications in various studies of quasi-one-dimensional strongly correlated systems 
(spin chains, itinerant electron systems, electron-phonon systems) are reviewed. 
}

\begin{document}

\maketitle

\section{Introduction}

The density-matrix renormalization group (DMRG) method~\cite{whit92b,whit93a}
was developed in 1992 to improve
the real-space renormalization group approach to
quantum lattice systems such as the Heisenberg and Hubbard models.
Since then density-matrix renormalization approaches have
been applied to a great variety of
problems in all fields of physics and even in quantum chemistry.
In this contribution I will review the calculation
of momentum- and frequency-resolved dynamical correlation functions
in low-dimensional strongly correlated systems using DMRG methods.
Numerous other extensions and applications of DMRG are discussed in various review 
articles~\cite{scho05,hall06} and books.~\cite{dmrg_book,greifswald} 
Additional information about DMRG
can be found at {\it http://www.dmrg.info}.

The outline of this contribution is as follows:
In the rest of this section I introduce the basic
DMRG algorithm for computing quantum states in
lattice models. In the next section I discuss
four DMRG methods for calculating dynamical correlation functions
(the Lanczos-vector method, the correction-vector method, the variational
method, and the time-evolution approach) and the techniques used
to obtain momentum-resolved spectra with DMRG.
In the last section I review important applications
of these methods to low-dimensional strongly correlated systems.

\subsection{DMRG and matrix-product states}

The key idea of DMRG is the renormalization of a quantum system 
using the information provided by a reduced density matrix rather than
an effective Hamiltonian as done in most other
renormalization group methods.  
Recently, the connection between DMRG and matrix-product states (MPS) has 
lead to significant progress.~\cite{mccu07}
DMRG is now considered to be the most efficient algorithm for optimizing
a variational MPS wavefunction.
The conceptual background of DMRG and MPS is discussed
in Ref.~\citen{Peschel} 
and the basic DMRG algorithms are presented in detail in 
several publications.~\cite{whit93a,dmrg_book,noac05,Jeckelmann}
Here I will summarize some basic features of the DMRG approach
which are necessary to understand its extension to the computation 
of dynamical correlation functions.
For this purpose I will use both the new MPS formalism 
and the traditional formulation in terms of 
blocks and superblocks. 

We consider a quantum lattice system with $N$ sites $n=1, \dots, N$.
Let $\{| s_n\rangle; s_n = 1, \dots , d_n\}$ denotes
a complete orthonormal basis of the Hilbert space for site $n$. 
(For instance, $\{ |\uparrow\rangle, |\downarrow\rangle \}$
for the spin-$\frac{1}{2}$ Heisenberg model.)
The tensor product of these bases yields a complete basis of the system
Hilbert space $\cal{H}$
\begin{equation}
\{ | \vec{s} = (s_1, \dots, s_N) \rangle = |s_1\rangle \otimes \dots 
\otimes |s_N\rangle \} .
\end{equation}
Any state $|\psi\rangle$ of $\cal{H}$ can be expanded in this basis
\begin{equation}
|\psi\rangle = \sum_{\vec{s}} c(\vec{s}) | \vec{s} \rangle.
\label{expansion}
\end{equation}
In the DMRG approach the coefficients
$c(\vec{s})$ take the form of a particular MPS  
\begin{equation}
c(\vec{s}) = \tens{A}_1(s_1) \dots \tens{A}_j(s_j) \tens{C}_j
\tens{B}_{j+1}(s_{j+1}) \dots \tens{B}_N(s_N) ,
\label{mps}
\end{equation}
where $\tens{C}_j$ is a $(a_{j} \times b_{j+1})$-matrix (i.e., with 
$a_{j}$ rows and $b_{j+1}$ columns). 
The $(a_{n-1} \times a_{n})$-matrices $\tens{A}_n(s_n)$ 
and the $(b_{n} \times b_{n+1})$-matrices $\tens{B}_n(s_n)$ 
fulfill the orthonormalization conditions
\begin{equation}
\sum_{s_n=1}^{d_n} \left ( \tens{A}_n(s_n) \right )^{\dagger} \tens{A}_n(s_n)  =  \tens{I}
\hspace{4mm} {\rm and} \hspace{4mm}
\sum_{s_n=1}^{d_n} \tens{B}_n(s_n) \left ( \tens{B}_n(s_n) \right )^{\dagger}  =  \tens{I} 
\label{constraints}
\end{equation}
($\tens{I}$ is the identity matrix) and the boundary conditions
$a_{0} = b_{N+1} = 1$.
Thus the square norm of $| \psi \rangle$ is given by
$\langle \psi | \psi \rangle = {\rm Tr} \ \tens{C}_j^{\dagger} \tens{C}_j$.

Any state $| \psi \rangle \in {\cal H}$ can be written
in the form~(\ref{mps}) using matrices with dimensions
$a_{j} = \prod_{n=1}^j d_n$ and $b_{j+1} = \prod_{n=j+1}^N d_n$.
However, this means that matrix
dimensions become exponentially large with increasing system size
(up to $2^{N/2}$ for a spin-$\frac{1}{2}$
model). Currently, a MPS is numerically tractable only if
all matrix dimensions are relatively small (up to a few thousands). 
A MPS with restricted matrix sizes ($a_{j} \leq \prod_{n=1}^j d_n$, 
$b_{j+1} \leq \prod_{n=j+1}^N d_n$)
can be considered as an approximation for states in $\cal{H}$.
In particular, it can be used as a variational ansatz for the ground
state of the system Hamiltonian $H$. Thus
the system energy becomes a 
function of the matrices $\tens{A}_n(s_n)$, $\tens{B}_n(s_n)$, and
$\tens{C}_j$
\begin{equation}
E=\frac{\langle \psi | H | \psi \rangle}{\langle \psi | \psi \rangle}
= E(\{\tens{A}_n(s_n)\},\{\tens{B}_n(s_n)\},\tens{C}_j) \ .
\label{energy}
\end{equation}
To determine the variational ground state
this function has to be minimized with respect to the variational
parameters $\tens{A}_n(s_n)$, $\tens{B}_n(s_n)$, and
$\tens{C}_j$ subject to the constraints~(\ref{constraints}). 
In the following subsection I will discuss the finite-system DMRG method,
which is the most efficient approach for carrying out this minimization.

\begin{figure}[tb]
	\centerline{\includegraphics[width=3cm]{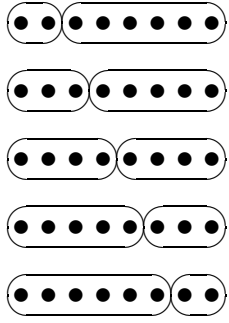}}
	\caption{Schematic representations of 
	the finite-system DMRG algorithm for a lattice with $N=8$ sites. 
	Solid circles are lattice sites and ovals
	are blocks. Going from top to bottom corresponds to
	iterations from $j=2$ to $j=N-2=6$ in a sweep from left to right
	while going from bottom to top corresponds to iterations
	form $j=6$ to $j=2$ in a sweep from right to left.  
	}
	\label{fig1}       
\end{figure}

Obviously, the MPS~(\ref{mps}) splits the lattice sites in two groups.
The sites $n=1, \dots, j$ make up a left block $L(j)$ and the 
sites $n=j+1, \dots, N$ constitute a right block $R(j+1)$, see fig.~\ref{fig1}.  
Using matrices $\tens{A}_n(s_n)$ and $\tens{B}_n(s_n)$ 
which satisfy the orthonormalization conditions~(\ref{constraints})
one can define a set of $a_{j}$ orthonormal states in the
Hilbert space associated with the left block 
\begin{equation}
\left | \phi^{L}_{\alpha} \right \rangle  =  
\sum_{s_1=1}^{d_1} \dots \sum_{s_j=1}^{d_j} \
\tens{A}_1(s_1) \dots \tens{A}_j(s_j) \
| s_1 \rangle \otimes \dots \otimes | s_j \rangle 
\end{equation}
($\alpha$ is the column index of the matrices $\tens{A}_j(s_j)$)
and a set of $b_{j+1}$ orthonormal states in the Hilbert space associated with the right block
\begin{eqnarray}
\left | \phi^{R}_{\beta} \right \rangle  =
\sum_{s_{j+1}=1}^{d_{j+1}} \dots \sum_{s_{N}=1}^{d_{N}} \
\tens{B}_{j+1}(s_{j+1}) \dots \tens{B}_N(s_N) \
| s_{j+1} \rangle \otimes \dots \otimes | s_{N} \rangle \ 
\end{eqnarray}
($\beta$ is the row index of the matrices $\tens{B}_{j+1}(s_{j+1})$).

These states span a subspace
of the Hilbert space associated with the left block and the right block, respectively.
Using these states one can build 
renormalized (i.e., approximate)
block representations of chosen dimension $a_j$ and $b_{j+1}$. 
Combining the left block $L(j)$ with the right block $R(j+1)$, we obtain the so-called 
superblock $\{ L(j) + R(j+1) \}$ which contains the sites $1$ to $N$.
The set of orthonormal tensor-product states 
\begin{equation}
\{
| \alpha \ \beta \rangle = 
| \phi^{L}_{\alpha} \rangle \otimes | \phi^{R}_{\beta} \rangle 
\}
\label{superblock_basis}
\end{equation}
spans a ($a_{j} b_{j+1}$)-dimensional subspace of the system Hilbert
space $\cal{H}$ and  
is called a superblock basis. 
A state represented by a MPS~(\ref{mps}) 
can be expanded in this basis
\begin{equation}
| \psi \rangle = \sum_{\alpha=1}^{a_{j}} \sum_{\beta=1}^{b_{j+1}}
[\tens{C}_j](\alpha,\beta) \ 
| \alpha \ \beta \rangle ,
\label{superblock_expansion}
\end{equation}
where $[\tens{C}_j](\alpha,\beta)$ denotes the matrix elements of $\tens{C}_j$,
(i.e., the elements of the matrix $\tens{C}_j$ are the components of the
state $| \psi \rangle$ in the superblock basis).

\subsection{Finite-system DMRG algorithm}
\label{sec:finite}

The finite-system algorithm allows us to determine the optimal superblock basis 
(i.e. the optimal matrices $\tens{A}_n(s_n)$ and $\tens{B}_n(s_n)$ with restricted
matrix dimensions)
to represent selected quantum states (the so-called target states).
It is the most versatile DMRG algorithm as it can readily be applied to almost any quantum 
lattice problem and has already been 
used to study spin, fermion, and boson systems in one and higher dimensions.  
It is also the most reliable DMRG algorithm as it always converges to the best possible MPS
representation~(\ref{mps}) for the target states. A detailed description
of this algorithm can be found in Ref.~\citen{Jeckelmann}.

In the finite-system algorithm the superblock structure $\{ L(j) + R(j+1) \}$
is moved iteratively by one site from $j=2$ to $j=N-2$ in a sweep
from left to right and from $j=N-2$ to $j=2$ in a sweep from right to left, 
see fig.~\ref{fig1}.
At each iteration we first determine the matrix representations of
quantum operators in the superblock basis~(\ref{superblock_basis}),
especially the Hamiltonian $H$.
Then the superblock representations  $\tens{C}_j$ of all target states are calculated,
in particular the ground state of the superblock Hamiltonian. 
In DMRG calculations for dynamical correlation functions additional states are 
targeted (see the next section).

Once the superblock representations $\tens{C}_{j}$ of the target states
have been calculated, a new basis 
which describes the target states as closely as possible
is constructed for the next superblock  with $j+1$
(left-to-right sweep) or $j-1$ (right-to-left sweep) substituted for $j$
in equ.~(\ref{mps}). 
As discussed in Ref.~\citen{Peschel} this can be done using the Schmidt
decomposition of $\tens{C}_{j}$ for a single target state.
More generally, for several target states the optimal approach 
consists in selecting the eigenvectors of reduced density matrices
with the highest eigenvalues.  
Therefore, if the DMRG calculation targets a state with a
vector representation $\tens{C}_{j}$ in the superblock basis, 
we calculate the reduced density matrix for the left block 
\begin{equation}
\rho(\alpha,\alpha') =  
\sum_{\beta} 
\left ( [\tens{C}_{j}](\alpha,\beta) \right )^*
[\tens{C}_{j}](\alpha',\beta) 
\label{density_matrix_left}
\end{equation}
or for the right block 
\begin{equation}
\rho(\beta,\beta') = 
\sum_{\alpha}
\left ( [\tens{C}_{j}](\alpha,\beta) \right )^*
[\tens{C}_{j}](\alpha,\beta')   .
\label{density_matrix_right}
\end{equation}
Reduced density matrices have eigenvalues $w_{\mu} \geq 0$
with $\sum_{\mu} w_{\mu} = 1$.
The $m$ eigenvectors with the largest eigenvalues $w_{\mu}$ are used to 
construct new block bases for the next iteration
while the other eigenvectors are discarded.
Thus, we can obtain a representation (basis) of chosen dimensions
$a_k, b_{k+1} \leq m$ ($k=j\pm 1$)
for the blocks constituting the next superblock. 

\begin{figure}[tb]
\centerline{  
\includegraphics[width=6cm]{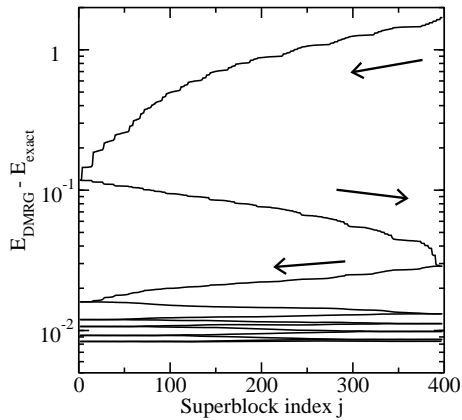}
}
\caption{Convergence of the ground state energy
calculated with the finite-system DMRG algorithm
using $m=20$ density-matrix eigenstates
as a function of the iterations
in a 400-site spin-$\frac{1}{2}$ Heisenberg chain.
Arrows show the sweep direction for the first three sweeps.
}
\label{fig2}       
\end{figure}

Iterations from one superblock to the next one are continued 
until the sweep is completed. 
Then we perform a sweep in the opposite direction.
The superblock basis (i.e., the matrices $\tens{A}_n(s_n)$ and
$\tens{B}_n(s_n)$) converges progressively to optimal values
for representing the target states as we perform
sweeps back and forth. For instance, in ground state
calculations, the variational energy~(\ref{energy}) 
decreases as
the sweeps are performed because of the
progressive optimization
of the variational MPS~(\ref{mps}) for the ground state.
Figure~\ref{fig2} illustrates this convergence  
for the total energy of a 400-site Heisenberg chain.
The matrix dimensions $a_n$, $b_n$ are chosen to be not greater
than $m=20$ (the maximal number of density-matrix eigenstates kept at
each iteration). 
The sweeps are repeated until the ground state energy remains
(almost) constant.  
In fig.~\ref{fig2} the DMRG energy converges to a value 
$E_{\rm DMRG}(m=20)$ which lies about $0.008$ above the exact result
for the 400-site Heisenberg chain as expected for a variational approach.
The error introduced by the restriction of the matrix dimensions
$a_n, b_n  \leq m$ is called a truncation error.

If we target $M > 1$ states, the density matrix is formed as the sum
\begin{equation}
\rho =  \sum_{s=1}^{M} \ c_s \rho_s
\label{targets}
\end{equation}
of the density matrices $\rho_s = | \psi_s \rangle \langle \psi_s |$ 
for each target state.
As a result the DMRG algorithm 
produces a superblock basis describing these $M$ states 
as accurately as possible.
Here the coefficients $c_s > 0$ are normalized weighting factors
($\sum_s c_s = 1$), which allow us to vary the influence
of each target state in the formation of the density matrix.
In most cases, however,
this approach is limited to a small number $M$ of 
targets (of the order of ten) 
because DMRG truncation errors grow rapidly with the number of 
targeted states. 

\begin{figure}[tb]
\centerline{
\includegraphics[width=6cm]{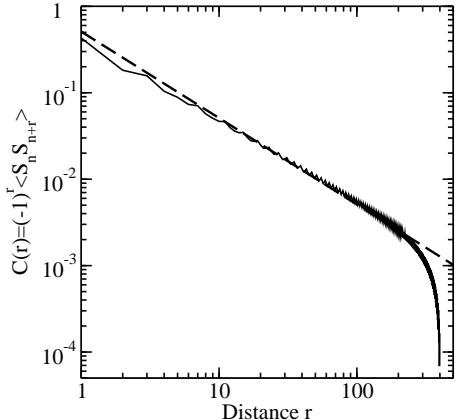}
}
\caption{
Staggered spin-spin correlation function 
$C(r)=(-1)^r \langle \vec{S}_n \vec{S}_{n+r} \rangle$
calculated using
the finite-system DMRG algorithm with $m=200$
in a 400-site spin-$\frac{1}{2}$ Heisenberg chain.
The dashed line shows $C(r)=0.51/r$ and is a guide for the eye.
}
\label{fig3}       
\end{figure}

Once convergence is achieved, observables can be calculated.
The finite-system algorithm
yields accurate results for expectation values $\langle \psi | {\cal O}|\psi \rangle$
of operators ${\cal O}$ with respect to target states $|\psi \rangle$.
For instance, in fig.~\ref{fig3} we show
the staggered spin-spin correlation function $C(r)=(-1)^r \langle \vec{S}_n \vec{S}_{n+r} \rangle$
obtained in the 400-site Heisenberg chain using up to $m=200$ density-matrix eigenstates.
For a distance up to $r \approx 100$
the staggered spin-spin correlation function $C(r)$ decreases approximately
as a power-law $1/r$ as expected but a deviation from this behavior
occurs for larger $r$ because of the chain edges.
Finite-size and chain-end effects are unavoidable and sometimes
troublesome features of the finite-size DMRG method.

\subsection{Truncation errors}
\label{sec:truncation}

There are various sources of numerical errors in the finite-system DMRG method.
First, errors can originate in the computation of superblock representations $\tens{C}_j$ 
for target states in a given superblock basis. 
These errors can always be made negligible although in computations of
dynamical correlation functions this can be very time consuming. 
Second, a superblock basis can be built using non-optimal matrices
$\tens{A}_n(s_n)$ and $\tens{B}_n(s_n)$ for given matrix dimensions
$a_n$ and $b_n$.  
If one performs enough sweeps through the lattice (up to several tens in hard cases), 
these errors can always be made smaller than truncation errors.

Truncation errors are the dominant source of inaccuracy in the finite-system
DMRG method and it is important to control them. They can be systematically 
reduced by increasing the matrix dimensions $a_n, b_n$.
A truncation error is introduced at every iteration
when a target state $|\psi\rangle$ which has been  
obtained in a superblock basis is approximated by a state 
$|\tilde{\psi}\rangle$ expanded in the next superblock basis.
To minimize the difference 
$S = | |\psi\rangle - |\tilde{\psi}\rangle |^2$ 
one has to select the eigenvectors 
with the highest eigenvalues $w_{\mu}$ from the reduced 
density-matrices~(\ref{density_matrix_left}) and~(\ref{density_matrix_right}). 
The minimum of $S$ is 
given by the weight $P$ of the discarded density-matrix eigenstates
and, assuming $w_1 \geq w_2 \geq \dots$, it can be written
$S_{\rm min} = P(a_{j}) = 1 - \sum_{\mu=1}^{a_{j}} w_{\mu}$ 
for the left block $L(j)$ and 
$S_{\rm min} = P(b_{j+1}) = 1 - \sum_{\mu=1}^{b_{j+1}} w_{\mu}$ for the right block $R(j+1)$.
Thus truncation errors are small when the discarded
weight is small. 
Experience shows that the accuracy of DMRG calculations
depends significantly on the system investigated because
the matrix-product state~(\ref{mps}) with restricted matrix sizes can be
a good or a poor approximation of targeted quantum states.  
For instance, the finite-system DMRG method yields excellent results
for the ground state of gapped one-dimensional systems but is less accurate 
for critical systems, excited states, or in higher dimensions.

\begin{figure}[tb]
\centerline{ 
\includegraphics[width=6cm]{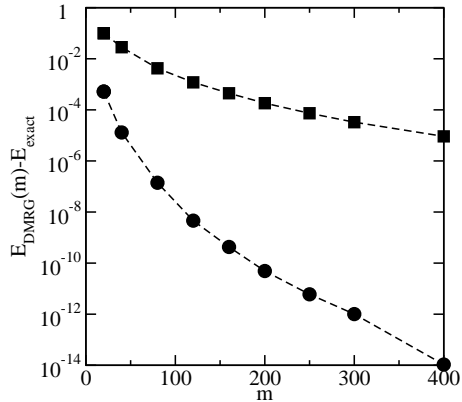}
}
\caption{Error in the ground state energy calculated
with the finite-system DMRG algorithm as a function
of the number $m$ of density-matrix eigenstates kept.
The system is the spin-$\frac{1}{2}$ Heisenberg model on a
one-dimensional 100-site lattice with open (circles)
and periodic (squares) boundary conditions. 
}
\label{fig4}       
\end{figure}

There are two established methods for 
choosing the matrix dimensions in a systematic way
in order to cope with truncation errors.
First, we can use matrix dimensions which are (almost) constant, $a_n, b_n \alt m$. 
In that case, the discarded weight is variable.
Second, the density-matrix eigenbasis can be truncated 
so that the discarded weight is approximately constant, $P(a_{n}), P(b_{n}) \alt P$.
In that case the number of density-matrix eigenstates kept 
(and so the matrix dimensions) is variable.
In both cases, physical quantities are calculated for several values
of $m$ or $P$ and their scaling is analyzed for increasing $m$ or decreasing $P$. 
As an example, in fig.~\ref{fig4} we show the truncation error in the ground state
energy $E_{\rm DMRG}(m)-E_{\rm exact}$ as a function of $m$
for a 100-site Heisenberg chain.  
For open boundary conditions (a favorable case) 
the error decreases very rapidly while
for periodic boundary conditions (a less favorable case)  
the error decreases more slowly as $m$ increases. 

The principal limitation of the DMRG method is the rapid increase of the computational 
effort with the system size in dimension larger than one and with the
range of the interactions.
Therefore, the majority of systems investigated with DMRG until now have been
(quasi-) one-dimensional systems with short-range interactions.
Theoretically, the computational cost 
is proportional to $N m^3$ for the number of operations
and to $N m^2$ for the memory (for a fixed number $m$ of density-matrix eigenstates are kept). 
As an example, 
the calculations shown in figs.~\ref{fig2} and~\ref{fig3}
took about 20 minutes on a 3 GHz Pentium 4 processor
and use less than to 300 MBytes of memory.
For more difficult problems with $m \approx 10^4$ or 
for studies of energy- and momentum-resolved
continuous spectra, the computational cost can reach
thousands of CPU hours and hundreds of GBytes of memory.

\section{Methods}

Calculating the dynamical correlation functions of strongly correlated systems
has been a long-standing problem of theoretical physics because
many experimental techniques probe these properties.
For instance, solid-state spectroscopy experiments, such as
optical absorption, photoemission, or nuclear magnetic resonance,
measure the dynamical correlations between an external time-dependent
perturbation and the response of electrons and phonons
in solids~\cite{kuzmany}.
Typically, the zero-temperature 
dynamic response of a quantum system at frequency $\omega$
(or equivalently energy $\hbar \omega$)
is given by a dynamical correlation function
(with $\hbar=1$)
\begin{equation}
G_{X}(\omega + \imag \eta,k) = - \frac{1}{\pi}
\left \langle \psi_0 \left | 
X_k^{\dag} \frac{1}{E_0+\omega + \imag \eta - H} X_k
\right |\psi_0 \right \rangle ,
\label{dynamic_CF}
\end{equation}
where $H$ is the time-independent Hamiltonian of the
system, $E_0$ and $|\psi_0 \rangle$ are its ground-state energy and
wavefunction, $X_k$ is a quantum operator corresponding to a physical
quantity characterized by a wavevector $k$ (or equivalently a momentum $\hbar k$),
and $X_k^{\dag}$ is the Hermitian conjugate of $X_k$. 
A small real number $\eta > 0$ is used to
shift the poles of the correlation function into the complex plane.

In general, we are interested in the imaginary part
of the correlation function 
\begin{equation}
\label{spectrum}
I_{X}(\omega + \imag\eta,k) = \mathrm{Im} \ G_{X}(\omega + \imag \eta,k) 
 =  \frac{1}{\pi} \left \langle \psi_0 \left | X_k^{\dag}
\frac{\eta}{(E_0+\omega -H)^2 + \eta^2} X_k \right |\psi_0 \right \rangle .
\end{equation}
for $\eta \rightarrow 0$.
For instance, the single-particle spectral function is the imaginary part
of the one-particle Green's function
\begin{equation}
\label{spectral}
A_{\sigma}(\omega \leq 0,k) = \lim_{\eta \rightarrow 0} 
\ I_{X}(-\omega + \imag \eta,k)
\end{equation}
for the operator $X_k = c_{k \sigma}$ which annihilates an electron with
spin $\sigma$ in the Bloch state with wavevector $k \in (-\pi,\pi]$.
This spectral function corresponds to the spectrum measured
in angle-resolved photoemission spectroscopy (ARPES) experiments.

Several approaches have been developed to calculate dynamical correlation
functions with DMRG. Here I will briefly present the four most relevant ones:
The Lanczos-vector method, the correction-vector method, the variational
method, and the time-evolution approach. Moreover, I will
discuss the techniques used to obtain momentum-resolved spectra.

\subsection{Lanczos-vector method}

The Lanczos-vector DMRG method~\cite{hall95,kueh99} 
combines DMRG with the Lanczos algorithm~\cite{gagl87}
to compute dynamical correlation functions.
Starting from the states
$|\phi_{-1} \rangle = 0$ and $|\phi_0\rangle = X_k |\psi_0\rangle$, 
the Lanczos algorithm
recursively generates a set of so-called Lanczos vectors:
\begin{equation}
|\phi_{n+1} \rangle  = H | \phi_n \rangle
- a_n |\phi_n \rangle - b^2_n |\phi_{n-1} \rangle  ,
\label{lanczos}
\end{equation}
where 
$a_n =  \langle \phi_n | H | \phi_n \rangle /
\langle \phi_n | \phi_n \rangle$ and
$b^2_{n+1} = \langle \phi_{n+1} | \phi_{n+1} \rangle/
\langle \phi_n | \phi_n \rangle$ 
for $n = 0, \dots , L-1$.
These Lanczos vectors span a Krylov subspace containing 
excited states contributing to the dynamical correlation 
function~(\ref{dynamic_CF}).
Calculating $L$ Lanczos vectors gives the first $2 L-1$ moments
of a spectrum and up to $L$
excited states contributing to it.
The dynamical correlation function is then given by the continued fraction expansion
\begin{equation}
- \pi G_{X}(z-E_0,k) = 
\frac{\langle \psi_0 | X_k^{\dagger} X_k|\psi_0\rangle}
{z-a_0-\frac{b_1^2}{z-a_1-\frac{b_2^2}{z-...}}}   .
\label{CF_expansion}
\end{equation}

This procedure (Lanczos iteration + continued fraction expansion)
has proved to be efficient and reliable
in the context of exact diagonalizations.~\cite{gagl87}
Within a DMRG calculation the Lanczos algorithm serves two purposes.
Firstly, it is used to compute the full dynamical spectrum using
representations of the relevant operators and Lanczos vectors
in a superblock basis (i.e., the matrices $\tens{C}_j$ representing
the states $| \phi_n \rangle$).
Secondly, the first few Lanczos vectors 
\{$n=0,\dots, M \leq L$\} are used as
target states in the reduced density matrix ~(\ref{targets}) 
in addition to the ground state $|\psi_0\rangle$.
Thus we can construct a superblock basis in which we can expand
both ground state and excited states
(i.e., we can find "optimal" matrices $\tens{A}_n(s_n)$ and $\tens{B}_n(s_n)$ for a
MPS representation of $|\psi_0\rangle$ and the states $| \phi_n \rangle$).
However, as DMRG truncation errors increase rapidly with
the number $M$ of target states,
only the first few Lanczos vectors (often only the first
one $|\phi_0\rangle$) are targeted in most applications.
As a result,
the density-matrix renormalization does not necessarily
converge to an optimal superblock basis for all excited
states contributing to a dynamical correlation function and
the calculated spectrum can be quite inaccurate. 
In particular, it often depends strongly on where the superblock 
is split in two blocks (i.e., the index $j$ in the MPS 
representation~(\ref{mps})).
Nevertheless, the Lanczos-vector  DMRG is a relatively
simple and quick method for calculating the dominant peaks
or the first few moments of
dynamical correlation functions within DMRG and
it has been used successfully
in several studies of low-dimensional strongly correlated 
systems (see Refs.~\citen{scho05,hall06}).
However, the shape of continuous spectra in large
systems can not be determined accurately with this  
method~\cite{kueh99}.

The DMRG method is usually implemented in real space because
its performance in momentum space are so poor that even ground state calculations
are very difficult.~\cite{Nishimoto}
However, if periodic boundary conditions are used in the real-space representation, 
wavevector-dependent operators $X_k$ can be expanded as
a function of local operators $X_j$, which act on a single site or bond only, 
using plane waves
\begin{equation}
X_{k} = \frac{1}{\sqrt{N}}
\sum^N_{j=1} \eul^{-\imag kj} X_{j}
\label{momentum}
\end{equation}
with wavevectors $k=2\pi z/N$ for integers $-N/2<z\leq N/2$.
For instance, the annihilation operators $c_{k \sigma}$ for electrons in
Bloch states, which are used in the definition of the
photoemission spectral functions $A(\omega,k)$, can be readily written
as a sum of annihilation operators $c_{j\sigma}$ 
for electrons localized on lattice sites 
\begin{equation}
c_{k\sigma} = \frac{1}{\sqrt{N}}
\sum^N_{j=1} \eul^{-\imag kj} c_{j\sigma} \ .
\end{equation}
If the Hamiltonian $H$ is translation invariant,
the Lanczos algorithm with the initial state $X_k | \psi_0 \rangle$
generates a Krylov space corresponding to states with a well-defined
momentum $\hbar (k+Q)$ where $\hbar Q$ is the momentum of the ground state.~\cite{hall95}
Therefore, it is possible obtain momentum-resolved correlation functions with
the Lanczos-vector DMRG method using periodic boundary conditions.  
The use of periodic boundary conditions is not too problematic for DMRG in this context
because the Lanczos-vector DMRG method is mostly applied to strongly correlated systems
on short one-dimensional lattices.

\subsection{Correction-vector method}

The correction vector~\cite{soos89} associated with the
dynamical correlation function $G_X(\omega + \imag \eta,k)$ is defined
by
\begin{equation}
\label{CV}
|\psi_X(\omega + \imag \eta,k) \rangle = \frac{1}{E_0+\omega +
\imag \eta - H}
| X_k \rangle \; ,
\end{equation}
where $| X_k \rangle = X_k | \psi_0 \rangle$ is identical to
the first Lanczos vector.
If the correction vector is known, the dynamical correlation
function can be calculated directly
\begin{equation}
G_X(\omega + \imag \eta,k) =
-\frac{1}{\pi} \langle X_k|\psi_X(\omega + \imag \eta,k) \rangle \; .
\label{CV_dynamic_CF}
\end{equation}
To calculate a correction vector 
an inhomogeneous linear equation system
\begin{equation}
(E_0+\omega + \imag \eta - H) | \psi \rangle = | X_k \rangle
\end{equation}
has to be solved for the unknown state $| \psi \rangle$.
Typically, the vector space dimension is very large and
the equation system is solved with the 
conjugate gradient method~\cite{numerical_recipes}
or other iterative methods~\cite{rama90}.
This approach can be extended to higher-order dynamic response 
functions such as third-order optical polarizabilities~\cite{pati99}. 

The correction-vector DMRG method~\cite{kueh99} consists in constructing
MPS representations~(\ref{mps}) of correction vectors~(\ref{CV}) and then in 
calculating the
corresponding dynamical correlation functions in a superblock
basis~(\ref{superblock_basis}) obtained this way.
The distinctive characteristic of the correction vector approach
is that a specific
quantum state~(\ref{CV}) yields the dynamical correlation function for
a given frequency $\omega$.  
In a DMRG calculation one can thus target a specific correction vector
and determine the dynamical correlation function for each frequency 
$\omega$ separately using a superblock basis
(i.e., matrices $\tens{A}_n(s_n)$ and $\tens{B}_n(s_n)$))
which have been optimized for that single excitation energy or a narrow
range around it. Therefore, truncation errors can be systematically reduced
using increasing matrix dimensions for MPS representations as done in
a ground state DMRG calculation. As a result,
the correction-vector DMRG method is much more accurate 
than the Lanczos-vector DMRG method, which uses the same superblock basis for all
frequencies.
However, the computational cost is also much higher
as the procedure has to be
repeated for many different frequencies to
obtain a complete dynamical spectrum.  
In practice, the correction-vector DMRG method allows one to
perform accurate calculations 
of complex or continuous spectra for all frequencies
in large lattices~\cite{kueh99,scho05,hall06}.

If the Hamiltonian $H$ is translation invariant, the correction 
vector~(\ref{CV}) belongs to the subspace of states with momentum
$\hbar (k+Q)$, where $Q$ is again the ground state wavevector. 
Thus as in the Lanczos-vector method
momentum-resolved correlation functions can be calculated with the
correction-vector DMRG using periodic
boundary conditions and momentum-dependent operators defined by equ.~(\ref{momentum}). 
However, since DMRG calculations are much more accurate (and thus
can be performed for much larger systems) with 
open boundary conditions than with  periodic boundary conditions
(see fig.~\ref{fig4}), it is desirable
to extend the definition of the momentum-resolved correlation functions
to the former case.   
Combining plane waves with filter functions in~(\ref{momentum})
is a possible approach to reduce boundary effects,
which has been
successfully used with the correction-vector DMRG method.~\cite{kueh99}

\subsection{Variational method}

The success of the correction-vector DMRG method shows that using specific target states
for each frequency is the right approach.   
This idea can be further improved using a variational formulation
of the problem~\cite{jeck02a,JeckelmannBenthien}.  Consider the functional
\begin{equation}
W_{X,k,\omega,\eta}(\psi)  = 
\langle \psi | (E_0+\omega-H)^2+\eta^2  | \psi \rangle
+ \eta \langle X_k | \psi \rangle + \eta \langle \psi | X_k \rangle \; .
\label{functional}
\end{equation}
For any $\eta \neq 0$ and a fixed frequency $\omega$ this functional
has a well-defined and non-degenerate minimum $| \psi_{\mathrm{min}} \rangle$.
This state is related to the correction vector~(\ref{CV}) by
\begin{equation}
(H-E_0-\omega+\imag\eta)| \psi_{\mathrm{min}} \rangle = 
\eta | \psi_X(\omega + \imag \eta,k) \rangle .
\label{CV2}
\end{equation}
The minimum is the
imaginary part of the dynamical correlation function
\begin{equation}
W_{X,k,\omega,\eta}(\psi_{\mathrm{min}}) =
-\pi\eta I_X(\omega + \imag \eta,k).
\label{minimum_spectrum}
\end{equation}
Thus the calculation of dynamical correlation functions can be formulated
as a minimization problem.

The DMRG method can be used to minimize a 
functional~(\ref{functional}) and thus to calculate the corresponding 
dynamical correlation function $G_X(\omega+\imag\eta,k)$.
This variational approach is called the dynamical DMRG (DDMRG) method.
The minimization of the functional is
easily integrated into the standard DMRG algorithm.
In the MPS formalism we want to minimize a function 
\begin{equation}
W(\{\tens{A}_n(s_n)\},\{\tens{B}_n(s_n)\},\tens{C}_j) = W_{X,k,\omega,\eta}(\psi)
\label{functional2}
\end{equation}
of the matrices $\tens{A}_n(s_n)$, $\tens{B}_n(s_n)$, and
$\tens{C}_j$ representing the state $| \psi \rangle$
similarly to the system energy~(\ref{energy}).
At every iteration in a sweep through the system lattice,
we calculate the minima of~(\ref{energy})
and~(\ref{functional2}) in the current superblock 
basis. In this way we obtain the superblock representations $\tens{C}_j$
of the states $|\psi_0\rangle$, $|X_k \rangle$, and
$|\psi_X(\omega+ \imag\eta,k)\rangle$, which are used as target~(\ref{targets})
of the density-matrix renormalization.
As in the correction-vector DMRG method we thus optimize the superblock basis
for a single frequency or a single narrow frequency range.  
Sweeps are repeated until the procedure has converged to the
minimum of~(\ref{functional2}).  
This minimum yields the imaginary part $I_X(\omega + \imag \eta,k)$ 
of the dynamical correlation 
function and the real part can be obtained as in the correction-vector
DMRG method.
To obtain a complete spectrum one has to repeat the calculation for 
numerous different frequencies $\omega$.
A more detailed description of the implementation of the DDMRG
algorithm can be found in Ref.~\citen{JeckelmannBenthien}. 

This variational formulation is completely equivalent to the
correction-vector method if we can calculate $|\psi_{\mathrm{min}}
\rangle$
and $|\psi_X(\omega + \imag \eta,k) \rangle$ exactly. 
However, if we can only calculate approximate states with an error
of the order $\varepsilon \ll 1$, 
the variational formulation~(\ref{minimum_spectrum})
gives the imaginary part $I_X(\omega + \imag \eta,k)$
with an accuracy of the order of $\varepsilon^2$, while the 
correction-vector approach~(\ref{CV_dynamic_CF}) yields results with
an error of the order of $\varepsilon$.
Consequently, the DDMRG method is 
more accurate than the correction-vector DMRG method for
the same computational effort or, 
equivalently, the DDMRG method is faster than the correction-vector
DMRG method for a given accuracy.
As found in a ground state DMRG calculations, numerical errors in DDMRG simulations
are dominated by truncation errors which can be systematically 
reduced using increasing matrix dimensions in the MPS representation~(\ref{mps}).
Numerous comparisons with exact analytical results and accurate numerical
simulations have demonstrated the unprecedented accuracy and reliability
of the DDMRG method in one-dimensional correlated 
systems of localized spins~\cite{nishimoto07}, 
of itinerant electrons~\cite{JeckelmannBenthien,jeck02a,jeck00,essler},
or of electrons coupled to phonons~\cite{jeck06}
and in 
quantum impurity problems~\cite{gebh03,nish04a}.
For one-dimensional strongly correlated electron systems such as 
the Hubbard model 
DDMRG allows for accurate calculations
of zero-temperature dynamical
properties for lattices with hundreds of sites and particles and
for any excitation energy.

\begin{figure}
\centerline{  
\includegraphics[width=8cm]{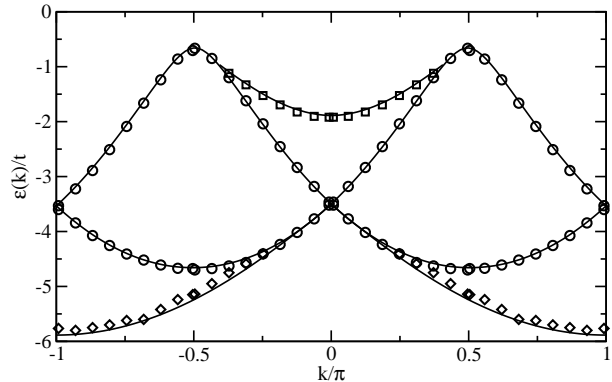}
}
\caption{The symbols show the dispersion of structures found in 
the single-particle spectral function of the one-dimensional
half-filled Hubbard model ($U=4t$) on a open-boundary 128-site chain 
using DDMRG and quasi-momenta:
Spinon branch (squares), holon branches (circles), and lower onset of the
spinon-holon continuum (diamonds).
Lines show the dispersion of corresponding excitation
branches calculated with the Bethe Ansatz for periodic boundary conditions.
}
\label{fig5}       
\end{figure}

To compute momentum-resolved spectra with DDMRG one can use periodic boundary conditions
and the operators~(\ref{momentum}) as done with the Lanczos-vector and correction-vector
DMRG methods. However, this approach often requires a prohibitive computational effort for
large systems ($N \agt 100$ for electronic systems) as 
DMRG performs much worse for periodic boundary conditions than for open boundary 
conditions (see fig.~\ref{fig4}).
Using open boundary conditions and plane waves with filter functions~\cite{kueh99} 
is also possible but this method is complicated and does not always yield good 
results~\cite{bent05b}.
A simple and efficient approach to compute momentum-resolved quantities
with DMRG consists in using open boundary conditions and 
operators defined by 
\begin{equation}
X_{k} = \sqrt{\frac{2}{N+1}} \sum^{N}_{j=1} \sin(kj) X_{j} 
\label{quasimomentum}
\end{equation}
with quasi-wavevectors $k=\pi z/(N+1)$ (quasi-momenta $\hbar k$)
for integers $ 1 \leq z \leq N$.
Both this expansion of $c_{k\sigma}$ and the conventional one~(\ref{momentum})
are equivalent in the thermodynamic limit
$N\rightarrow \infty$.
Numerous tests  have shown that
both approaches are also consistent in the entire Brillouin
zone for finite systems~\cite{JeckelmannBenthien,bent04,bent05b}. 
For instance, in Fig.~\ref{fig5}  
we compare the dispersion of excitations
in the single-particle spectral function of the one-dimensional Hubbard model
at half filling for $U=4t$. 
The agreement is excellent and allows us to identify the dominant structures, 
such as the spinon branch, two holon branches and the lower onset of the
spinon-holon continuum.~\cite{JeckelmannBenthien}
Therefore, the quasi-momenta~(\ref{quasimomentum})
can be used to investigate momentum-dependent quantities
such as spectral functions $A(\omega,k)$.

\subsection{Finite-size scaling}
\label{sec:scaling}

A DDMRG calculation is always
performed for a finite parameter $\eta > 0$ and
the obtained spectrum $I(\omega + \imag \eta)$
is equal to the convolution
of the true spectrum $I(\omega)$
with a Lorentzian distribution of width $\eta$
\begin{equation}
I(\omega + \imag \eta) =   
\int_{-\infty}^{+\infty} d\omega' I_{X}(\omega')
\frac{1}{\pi}\frac{\eta}{(\omega-\omega')^2+\eta^2} \; .
\label{convolution}
\end{equation}
Therefore, DDMRG spectra are always artificially broadened.
In particular, the broadening hides the discreteness of the spectrum
in finite-size systems. 
In the thermodynamic limit $N \rightarrow \infty$, 
a spectrum $I(\omega)$ may include continuous structures.
It is necessary to perform several calculations for
various $\eta$ to determine $I(\omega)$ accurately.
In the
thermodynamic limit, one has to calculate
\begin{equation}
I(\omega) =  \lim_{\eta \rightarrow 0} \lim_{N \rightarrow \infty}
I(\omega + \imag \eta)  .
\label{inflim}
\end{equation}
Computing both limits from numerical results is computationally expensive
and leads to large extrapolation errors.
A better approach is to use a broadening $\eta(N) >0$
which decreases with increasing $N$ and vanishes in the
thermodynamic limit~\cite{jeck02a}
\begin{equation}
I(\omega)  =  \lim_{N \rightarrow \infty} I(\omega + \imag \eta(N)) .
\label{inflim2}
\end{equation}
The function $\eta(N)$ depends naturally on the specific problem
studied  and
can also vary for each frequency $\omega$ considered.
For one-dimensional
correlated electron systems
one finds empirically that the optimal scaling is
\begin{equation}
\eta(N)  = \frac{c}{N} ,
\label{etacondition}
\end{equation}
where the constant $c$ is comparable 
to the effective
band width of the excitations contributing to
the spectrum around $\omega$. 
Thus features of some infinite-system spectra can be
determined accurately from DDMRG data for
finite systems.~\cite{JeckelmannBenthien,jeck02a,jeckelmann03,nishimoto07}
using a a size-dependent broadening $\eta(N)$.
It should however be noted that the scaling~(\ref{etacondition}) does
not hold for all systems. In particular, it does not seem appropriate for
electron-phonon systems such as the Holstein model.~\cite{jeck06}

A good approximation for a continuous infinite-system spectrum
can sometimes be obtained by deconvolution of the DDMRG data
for dynamical correlation functions.
A deconvolution consists in solving  	
the convolution equation~({\ref{convolution}) numerically
for an unknown smooth function $I(\omega')$
using DDMRG data for a finite system on
the left-hand side.
Performing such deconvolution is a ill-conditioned inverse problem,
which requires some assumptions on the spectrum properties such as
a finite width, a piecewise smoothness, and positive-semidefinite values. 
Typically the accuracy of deconvolved DDMRG spectra is unknown but comparisons with
exact results have shown that they are often accurate.
Excellent agreement has been achieved with exact results for the density 
of states of quantum impurities~\cite{gebh03,nish04a,raas05}
and the optical conductivity of one-dimensional Mott insulator.~\cite{jeck06}.
As an example, fig.~\ref{fig6} shows the DDMRG data ($\eta=0.1t$)
and the result of the deconvolution for the single-particle spectral function 
$A(\omega,k)$ of the spinless Holstein model on a half-filled 8-site ring.
The result of the deconvolution agrees well with the spectral function
obtained using exact diagonalization techniques and the
kernel polynomial method.~\cite{jeck06} In particular, the width of the
spectrum, which is difficult to estimate using the broadened DDMRG spectrum,
can be easily determined from the deconvolved spectrum.
A detailed discussion of deconvolution techniques for spectra
calculated with the DDMRG method or the correction-vector DMRG 
can be found in Ref.~\citen{raas05}.

\begin{figure}[tbh]
\centerline{  
\includegraphics[height=7cm]{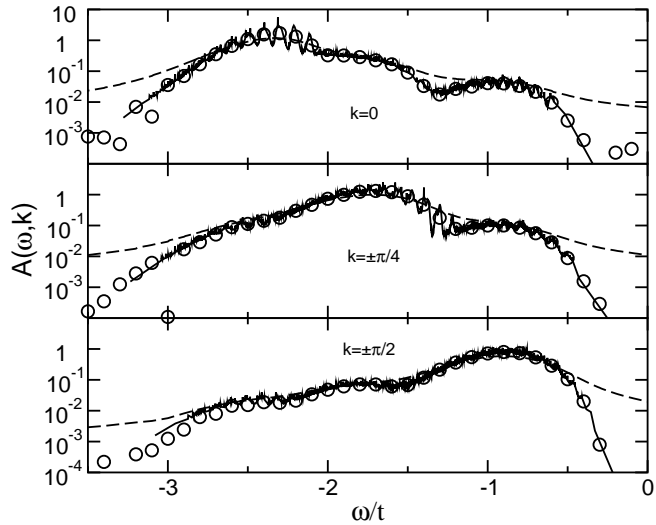}
}
\caption{Single-particle spectral function $A(\omega,k)$ of the 
spinless-fermion
Holstein model on a half-filled 8-site ring in the Peierls insulating phase. 
The dashed line shows the DDMRG spectrum with a
broadening $\eta=0.1t$. The result of the deconvolution is shown by circles.
The solid line shows exact diagonalization results for comparison.
Note the logarithmic scale of the vertical axis.
}
\label{fig6}       
\end{figure}

With DDMRG computing the spectrum~(\ref{spectrum}) for a single point
in the $(\omega,k)$ space is about as expensive as a ground state DMRG calculation.
In particular, the necessary CPU time scales linearly with the system size $N$.
However, to describe a spectrum which is continuous
in $\omega$ (for a fixed wavevector $k$) we have to calculate~(\ref{spectrum}) 
for many different frequencies $\omega$ with a separation $\Delta \omega \alt \eta$. 
If the broadening $\eta$ is scaled as~(\ref{etacondition}) when the system size $N$ increases,
the number of required frequencies increases linearly with $N$ (assuming a finite spectrum band width).
Thus the computational effort scales as $N^2$ if one calculates the full spectrum (as a function
of $\omega$) for a fixed $k$.
The number of different wavevectors $k$ in~(\ref{momentum}) or~(\ref{quasimomentum}) is $N$.
Thus the total computational effort for calculating the full spectrum~(\ref{spectrum})
for all wavevectors $k$ is proportional to $N^3$.
Fortunately, as DDMRG calculations for different points $(\omega,k)$
can be performed independently, this approach can be easily parallelized. The parallelization
of a single ground state DMRG calculation or of a DDMRG calculation for a single $(\omega,k)$-point
is also possible but more difficult.~\cite{Georg}

\subsection{Time-evolution approach}

A major advance in the DMRG method in recent years has been 
the development of several techniques
for the simulation of the real-time evolution in one-dimensional strongly correlated 
systems.~\cite{scho05,scho06,Noack}  
These techniques allow us to integrate the Schr\"{o}dinger equation
\begin{equation}
i\hbar \frac{d}{dt} |\phi(t)\rangle  = H |\phi(t)\rangle
\end{equation}
starting from an initial state $|\phi(t_0)\rangle$.
The state $|\phi(t)\rangle$ is calculated for discrete time steps $\tau$
at which it is represented by a MPS. This MPS has the form~(\ref{mps}) 
in some algorithms but other representations are also used.  
The computational effort scales linearly with $T/\tau$.
As in the previously discussed DMRG methods, 
truncation errors are the main source of inaccuracies
in time-dependent DMRG simulations. 
They accumulate exponentially in time and lead
to a runaway time $T$ beyond which time-dependent DMRG simulations break down. 
For time $t_0 \leq t \alt T$, however, the best time-dependent DMRG methods yield
results which seem to be as accurate as in conventional DMRG simulations. 

The dynamical correlation function $G_{X}(\omega + \imag \eta,k)$ defined in
equ.~(\ref{dynamic_CF}) 
is the Laplace transform (up to a prefactor)
of the time-resolved correlation function 
\begin{equation}
\label{time_CF}
G_{X}(t\geq 0) = \langle \psi_0 |
X_k^{\dag}(t) X_k(0) | \psi_0 \rangle = \exp\left( \frac{iE_0t}{\hbar}\right ) 
\langle\phi(0)|\phi(t)\rangle ,
\end{equation}
where $X_k(t)$ is the Heisenberg representation of the operator $X_k$
and the initial condition is $| \phi(t_0 = 0) \rangle = X_k|\psi_0\rangle$.
Thus one can obtain $G_{X}(\omega + \imag \eta,k)$ with a resolution
$\Delta \omega \sim \pi/T$ through a Laplace transformation
of the time-resolved DMRG data for~(\ref{time_CF}) with $\eta \propto 1/T$ 
(or a Fourier transformation with a windowing function of width $\propto T$).
The time-resolved DMRG data can also be extrapolated for large times using
linear prediction techniques in order to enhance the frequency resolution.~\cite{white08}   
However, the discrete time steps $\tau$ in the time-dependent DMRG simulations
lead to a high-frequency cut-off $|\omega| \alt \Omega = \pi/\tau$ in the spectrum of 
$G_{X}(\omega + \imag \eta,k)$. 
Therefore, the time-dependent DMRG approach is a priori more efficient
than frequency-approaches such as DDMRG for calculating a spectrum
over a large frequency range at low resolution (small $T$ and $\tau$) while
frequency-approaches should perform better when computing a spectral function
with high-resolution over a short frequency interval (small $\Omega$ and $\Delta \omega$).
A direct comparison of the time-dependent DMRG and DDMRG methods has not been
carried out yet, so that it is not clear how the performance of both methods
differs in practice.   

\section{Applications}

The DMRG methods discussed in the previous section have been successfully applied
to the study of dynamical correlations, dynamical response functions, and excitation spectra
in a great variety of one-dimensional strongly correlated quantum systems
and quantum impurity problems. In this section I will review some of the most important
applications and results obtained so far.

\subsection{Spin chains}

DMRG methods for dynamical properties have been systematically used to 
investigate the dynamical spin structure factor and excitation spectrum
of quantum spin chains.  
The dynamical structure factor $S(\omega,k)$ corresponds to an
energy- and momentum-resolved
spin-spin correlation function~(\ref{spectrum})
with $X_k = S^z_k$ or $S^{\pm}_k$. 
As several exact results
are available for these systems, they also offer a good opportunity
for testing the accuracy of numerical methods such as DMRG.

In the original work describing the Lanczos-vector DMRG approach~\cite{hall95}
Hallberg has illustrated the method with an investigation of the 
dynamical structure factor of a $S=1/2$ isotropic 
Heisenberg chain with up to 72 spins.  The dispersion relation of 
the lowest excitation has been determined from the DMRG data for the spectrum
and the validity of the Lanczos-vector DMRG approach has been demonstrated by comparison 
with the exact dispersion from the Bethe Ansatz solution.

In the paper introducing their implementation of the correction-vector
DMRG method~\cite{kueh99}  K\"{u}hner and White have investigated
the dynamical structure factor of the $S=1$ and $S=1/2$ Heisenberg chains 
with up to 320 sites using both Lanczos-vector and correction-vector DMRG methods. 
They have shown that the correction-vector approach is more accurate and more efficient
than the Lanczos-vector approach when combined with DMRG and applied to large systems. 
In the $S=1$ Heisenberg chain the weight and energy
of the single magnon excitation has been determined. The DMRG dispersion
agrees perfectly with exact diagonalization and quantum Monte Carlo (QMC) results. 
In the $S=1/2$ system K\"{u}hner and White have confirmed that the lowest excitation
calculated with the Lanczos-vector DMRG method
and the dispersion of the continuum onset calculated from the correction-vector DMRG data 
agree well with the Bethe Ansatz solution. Moreover, 
they have demonstrated that the correction-vector DMRG method can be used to study a continuous
spectral function of $\omega$ by computing the shape of the continuum in $S(\omega,k)$  
at $k=\pi$.

Nishimoto and Arikawa~\cite{nishimoto07} have studied the dynamical structure factor
of $S=1/2$ Heisenberg chains with uniform and  staggered magnetic fields
using DDMRG. They have found that their DDMRG results agree qualitatively with spectral line shapes 
derived from the Bethe Ansatz solution. At low-frequency, where these spectral line shapes are 
exact, they have obtained a satisfactory quantitative agreement with DDMRG data.

Recently, the time-dependent DMRG has been used to calculate the
dynamical structure factor of the $S=1/2$ $xxz$ spin chain with
up to 400 sites.~\cite{Pereira} The obtained DMRG data are in
excellent agreement with formula for the singularities in $S(\omega,k)$
and thus confirm the validity of these analytical predictions.
Moreover, the dynamical structure factor of the $S=1$
antiferromagnetic Heisenberg chain with up to 400 sites has been determined
using the time-dependent DMRG supplemented by a linear prediction method 
for extrapolating time-resolved data to longer times.~\cite{white08} 
This approach has yield impressively accurate spectral functions, which
allow for a study of fine details of the spectrum properties, in particular
the region where the single-magnon excitation meets the two-magnon continuum.

Among other applications of DMRG to quantum spin chains we mention
a study~\cite{Yu} of the dynamical structure factor in the one-dimensional 
spin-orbital model in a magnetic field, which has presented the first
calculation of full spectra in the $(\omega,k)$ space using the Lanczos-vector
and correction-vector DMRG methods;
an investigation~\cite{Friedrich} of edge singularities in the
$S=1$ Heisenberg chain in a strong external magnetic field exceeding the
Haldane gap using a MPS generalization of the correction-vector method with 
a separate MPS representation for each target state; and finally a
calculation~\cite{Okunishi} of the dispersion of the lowest excitation 
in the dynamical structure 
factor of the $S=1$ bilinear-biquadratic chain with up to 240 sites
using the Lanczos-vector DMRG method.

\subsection{Electronic systems}

DMRG methods for dynamical properties have been used to investigate various
excitations and dynamical response functions in one-dimensional itinerant electron systems
such as the Hubbard model and its extensions. These calculations are significantly
more difficult than those for spin chains and in exhaustive
calculations of momentum- and energy-resolved
correlation functions (i.e., for all relevant values of $\omega$ and $k$) system sizes rarely
exceed $N=100$ sites. 

The first applications (and still among the most frequent ones) have been
studies of the linear optical absorption and optically excited states, especially excitons,
in quasi-one-dimensional Mott or Mott-Peierls 
insulators such as conjugated polymers or cuprate chains
(for instance, see 
Refs.~\citen{pati99,jeck02a,jeck00,jeckelmann03,nishi03,Matsueda01,Matsueda02,Benthien3}).
The optical absorption is proportional to the dynamical current-current or dipole-dipole
correlation function but optically-allowed excitations have a momentum $k \rightarrow 0$
(relative to the ground state). Therefore, a momentum-resolved DMRG method
is not necessary for these applications and I will not discuss them in more detail. 

DMRG methods have also been employed to investigate the spectral function of
quantum impurity problems such as the single impurity Anderson 
model.~\cite{raas05,nish04a,raas04,Nishimoto06}
Moreover they have been successfully used as impurity solver in the framework
of the dynamical mean-field theory (DMFT) for the Hubbard model in the limit of
high dimensions.~\cite{gebh03,Garcia,Karski}
In both types of application it has been found that DMRG methods are useful complement to
existing ones (such as QMC simulations and numerical renormalization group).
For instance, DMRG methods can determine the high-frequency part of the
zero-temperature spectral function with high resolution, especially the Hubbard 
satellites.~\cite{gebh03,nish04a,raas04,Karski}
The impurity spectral function is a local dynamical correlation functions,
not a momentum-resolved one. Thus I will not discuss
this type of calculation further.

A first momentum-resolved DMRG calculation for dynamical correlations in 
electronic systems has been performed to explain the resonant inelastic
x-ray scattering (RIXS) spectrum of the quasi-one-dimensional compound SrCuO$_2$.~\cite{Kim}
In first approximation a cuprate chain can be described by a 
one-dimensional extended 
Hubbard model (EHM) with nearest-neighbor repulsion
at half filling. In Ref.~\citen{Kim}  the dynamical charge
structure factor $N(\omega,k)$ of this model has been calculated using DDMRG
and quasi-momenta~(\ref{quasimomentum}). 
(The dynamical charge structure factor is the dynamical correlation
 function~(\ref{dynamic_CF}) with the operator $X_k = n_k$.)
This investigation has shown that the main features of the RIXS spectrum
(dispersion of the continuum onset and of the intensity maximum),
the low-energy optical absorption, and the spin excitation band width
can be explained by the EHM using a single set of model parameters.

This first DDMRG study has been recently extended by a comprehensive
investigation 
of the spin and charge dynamics of the one-dimensional EHM
at half-filling.~\cite{Benthien} 
It confirms that
the low-energy dynamics of the cuprate chains SrCuO$_2$ can be 
described by the EHM with a single set of model parameters
and that this system is a 
quasi-one-dimensional Mott insulator.
In particular, we can understand the results of optical absorption
(dynamical current-current correlations), neutron scattering
(dynamical spin structure factor), RIXS 
(dynamical charge structure factor), and ARPES  
(one-particle spectral function) experiments within this framework.

In Ref.~\citen{Benthien}  
a similar conclusion has been drawn from partial results 
for the parent cuprate compound Sr$_2$CuO$_3$ using slightly 
different model parameters.
In a very recent work~\cite{Matsueda3} the effects of phonons on the
linear optical absorption and possible excitons 
have been investigated using an extended Hubbard-Holstein
model and the correction-vector DMRG method. The results suggest
that phonons are necessary to explain the linear absorption
spectrum of Sr$_2$CuO$_3$.  

The effects of phonons on the ARPES spectrum of one-dimensional
Mott insulators have also been investigated using the Holstein-Hubbard
model and DMRG methods.~\cite{Matsueda2}
It has been found  that the main features,
especially the spin-charge separation, are robust with respect to
realistic electron-phonon coupling and that the experimental ARPES results
for SrCuO$_2$
are consistent with the theoretical DMRG results.
A comparison of quantum Monte Carlo simulations for finite temperature
with zero-temperature DMRG data has confirmed that the main features
of the spectral function in the half-filled Hubbard model
are not modified qualitatively at finite but low
temperature.~\cite{Matsueda1}

The DDMRG method as also been used for an extensive study of various
extended Hubbard models with nearest-neighbor repulsion and hopping terms
at half filling.~\cite{Hoinkis} 
This has been motivated by the unusual and unexplained  dispersion
observed in one direction in the ARPES spectrum of the 
compound TiOCl. Unfortunately, the considered interactions do not change
the single-particle spectral function qualitatively as long as 
the ground state remains a Mott insulator and
no satisfactory explanation for 
the ARPES results has been found so far. It is likely that
a realistic description of this material requires a multi-band model
and the consideration of multiple chains,
which is beyond the present capability of DMRG methods.

A detailed study of the charge and spin dynamics has also
been carried out for the one-dimensional quarter filled Hubbard model 
with next-nearest neighbor hopping integrals using the DDMRG method.~\cite{Ejima}
This model is believed to be relevant for some quasi-one-dimensional
organic conductors of the Bechgaard salt family (TMTSF)$_2$X.
DDMRG results for the dynamical charge and spin structure factors
support the spin-triplet pairing mechanism  for
the superconducting phase of these materials.

One of the most demanding applications of the momentum-resolved DMRG approach
has been the study of the single-particle spectral function in the one-dimensional
Hubbard model away from half filling.~\cite{bent04}
This system is a Luttinger liquid with two gapless excitation modes
corresponding to collective spin (spinon) and charge (holon)  excitations, respectively. 
An accurate MPS representation of the system excited states is quite difficult
in such a case.  Nevertheless this study has been successfully completed
using the DDMRG method and the quasi-momentum technique~(\ref{quasimomentum}).
It shows that the dynamic separation of spin and charge
predicted by field theoretical methods in the asymptotic limit $\omega \rightarrow$
can be observed at finite excitation energy $\hbar\omega$ in the single-particle
spectral function of the Hubbard model. For less than half filling separate
spinon and holon branches are clearly visible as dispersive peaks (maxima)
in the spectral weight distribution, see fig.~\ref{fig7},
while for more than half filling only the holon branch corresponds
to dispersive peaks and the spinon branch gives the low-energy onset of the spectrum.

\begin{figure}[tb]
\centerline{ 
\includegraphics[width=10cm]{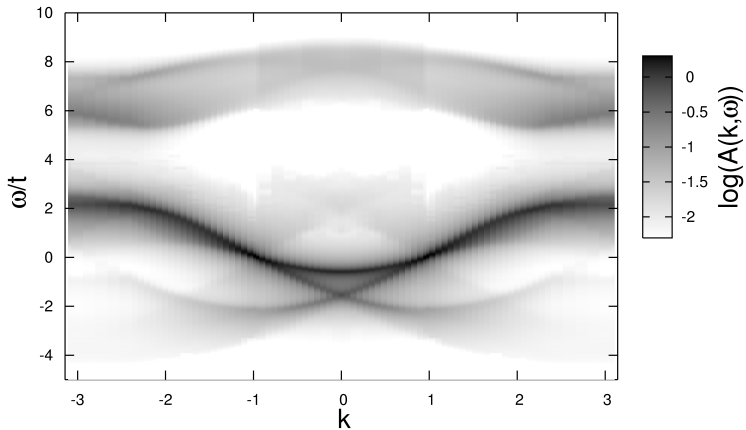}
}
\caption{Density plot of the single-particle spectral function of the 
one-dimensional Hubbard model for $U=4.9t$ and less than half filling ($\rho = 0.6$ electrons per site). 
The spectral function has been calculated on a 90-site open chain using DDMRG
with $\eta=0.1t$ and quasi-momenta. The ARPES spectrum corresponds to $\omega < 0$
and the inverse ARPES spectrum to $\omega > 0$. The spectral function for more than half filling
($\rho = 1.4$) is obtained through the transformation $(\omega,k) \rightarrow (-\omega,k+\pi \mod 2\pi)$.}
\label{fig7}       
\end{figure}

This DDMRG study of the Hubbard
model spectral function confirms that dispersive features
observed in the ARPES spectrum of the quasi-one-dimensional conductor TTF-TCNQ are
the signature of the spin-charge separation in one-dimensional strongly correlated
electron systems.  
The accuracy of the DDMRG results
has been demonstrated by a comparison with the exact dispersion 
of excitations obtained from the Bethe Ansatz solution (as shown in 
fig.~\ref{fig5} for the half-filled band case) and later confirmed by
QMC simulations.~\cite{Abendschein,Bulut}
Finite-temperature effects and the role of phonons have also been
investigated using DMRG methods.~\cite{Bulut,Ning}   

\subsection{Electron-phonon systems}

Electron-phonon systems are a challenge for DMRG simulations because
the Hilbert space of a single phonon site is infinitely large. To deal
with this problem a density-matrix renormalization approach has been developed
to find an optimal finite-dimensional basis for phonon (or more generally
boson) sites.~\cite{jeck06} This optimal phonon basis technique can be combined
with the Lanczos-vector approach to compute momentum- and energy-resolved
dynamical correlation functions in electron-phonon models.
However, as the electronic degrees of freedom ae not renormalized using DMRG
in this approach but treated exactly,
its applicability is restricted to small system sizes.
Nevertheless, the method has been demonstrated on the single-particle spectral function
and the optical conductivity of the Holstein model at various band fillings.~\cite{Zhang}
Combined with cluster perturbation theory it allows us to obtain approximate single-particle
spectral functions with a higher resolution in $k$-space for infinite systems.~\cite{Ning}
This has been used to investigate the effect of phonons of the 
ARPES spectrum of TTF-TCNQ.

The combination of optimal boson basis and Lanczos algorithm has also been used
to investigate the dynamical susceptibility of a dissipative two-state system 
(a spin-boson model).~\cite{Nishiyama}
The obtained results agree with those of QMC simulations.

To calculate the dynamical correlations of large electron-phonon systems
one can treat both electron and phonon degrees of freedom with DMRG.~\cite{jeck06}
For instance, this approach has been used to compute the spectral functions of spin-polarized
electrons (spinless fermions) in the Holstein model, which are shown in fig.~\ref{fig5}.
This approach has also been employed to investigate the single-particle spectral 
function~\cite{Bulut,Matsueda2}
and the linear optical absorption~\cite{Matsueda3} 
in extended Holstein-Hubbard models with up to 20 sites.
These studies have shown that, 
for model parameters representing the Mott insulator SrCuO$_2$ or the
organic conductor TTF-TCNQ,
the electron-phonon coupling does not influence
$A(\omega,k)$ over the energy range observed in ARPES experiments.

\subsection{Cold gases in optical lattices}

One of the first applications of the correction-vector DMRG method has been the
calculation of the ac conductivity in the superfluid phase of the
one-dimensional Bose-Hubbard model for correlated bosons in a lattice.~\cite{Kuehner}
The advent of ultracold bosonic atom gases in optical lattices
has considerably increased the interest in the dynamics of these systems but
DMRG calculations for momentum- and energy-resolved dynamical correlation
functions remain scarce.
Recently, the spectral function $A(\omega,k)$ 
has been calculated in a two-component one-dimensional Bose-Hubbard model
using a correction-vector MPS method, which improves on the correction-vector DMRG
method.~\cite{Kleine}
Although the model considered describes cold atomic gases with two hyperfine species
in a quasi-one-dimensional optical lattice, a comparison with experiment is not possible 
because momentum- and energy-resolved spectral functions can not be measured
in cold atomic gases with the presently available techniques. Therefore,
for these systems
it is currently more interesting to investigate time-resolved quantities using one
of the time-dependent DMRG methods.

\section{Conclusion}

DMRG methods allow us to calculate the momentum- and energy-resolved
dynamical correlation functions of low-dimensional correlated systems 
on large lattices with several hundreds of sites. 
The accuracy of these DMRG calculations has been demonstrated by numerous comparisons with
exact results and numerical data obtained with other methods.  
The capability and versatility of DMRG methods
are illustrated by the broad range of applications summarized in the previous
section. The main drawback of this approach is 
the limitation to one-dimensional systems and quantum impurity systems
and to zero temperature.
An advantage of the DMRG approach over other numerical techniques
is that it allows for the simulation of
systems large enough to obtain information on the spectrum in
the thermodynamic limit. 
In summary, DMRG methods provides a powerful and versatile 
approach for investigating the dynamical properties
in low-dimensional strongly correlated quantum systems.


%

\end{document}